\documentclass{svproc}
\usepackage{url}
\usepackage{wrapfig}
\usepackage{relsize}
\usepackage{graphicx}
\usepackage{xcolor}
\usepackage[sorting=none]{biblatex}
\usepackage{tabularx}
    \newcolumntype{L}{>{\centering\arraybackslash}X}
\usepackage{array,multirow}
\usepackage{hyperref}
\usepackage{subfig}
\usepackage{inputenc}
\usepackage{float}
\usepackage{bm}
\usepackage{amsmath}
\usepackage{setspace}
\begin{document}
\mainmatter              
\title{COVIDScholar: An automated COVID-19 research aggregation and analysis platform}
\titlerunning{COVIDScholar}  
%
\author{Amalie Trewartha\inst{1} \and John Dagdelen\inst{1,2} \and
Haoyan Huo\inst{1,2} \and Kevin Cruse\inst{1,2} \and Zheren Wang\inst{1,2} \and Tanjin He\inst{1,2} \and
Akshay Subramanian \inst{3}\and 
Yuxing Fei \inst{4} \and Benjamin Justus \inst{2} \and 
Kristin Persson\inst{1,2} \and Gerbrand Ceder\inst{1,2}}
\authorrunning{Amalie Trewartha et al.} 
\institute{
Materials Sciences Division, Lawrence Berkeley National Laboratory, Berkeley, CA 94720, USA
\and
Department of Materials Science \& Engineering, University of California, Berkeley, Berkeley, CA 94720, USA
\and
Indian Institute of Technology Roorkee, Roorkee, Uttarakhand 247667, India
\and
Wuhan University, Wuhan, Hubei 430072, China}

\maketitle              
\begin{abstract}

    The ongoing COVID-19 pandemic has had far-reaching effects throughout society, and science is no exception. The scale, speed, and breadth of the scientific community's COVID-19 response has lead to the emergence of new research literature on a remarkable scale --- as of October 2020, over 81,000 COVID-19 related scientific papers have been released, at a rate of over 250 per day. This has created a challenge to traditional methods of engagement with the research literature; the volume of new research is far beyond the ability of any human to read, and the urgency of response has lead to an increasingly prominent role for pre-print servers and a diffusion of relevant research across sources. These factors have created a need for new tools to change the way scientific literature is disseminated.
    
    COVIDScholar is a knowledge portal designed with the unique needs of the COVID-19 research community in mind, utilizing NLP to aid researchers in synthesizing the information spread across thousands of emergent research articles, patents, and clinical trials into actionable insights and new knowledge. The search interface for this corpus, \mbox{\url{https://covidscholar.org}}, now serves over 2000 unique users weekly.
    
    We present also an analysis of trends in COVID-19 research over the course of 2020.

\end{abstract}

\section{Introduction}

The scientific community has responded to the COVID-19 pandemic with unprecedented speed, and as a result an enormous amount of research literature is rapidly emerging, at a rate of over 250 papers a day \cite{stats}. The urgency and volume of emerging research has caused pre-prints to take a prominent role in lieu of traditional journals, leading to widespread usage of pre-print servers for the first time in many fields, most prominently biomedical sciences\cite{10.1371/journal.pmed.1002549}\cite{Fraser2020.05.22.111294}. While this allows new research to be disseminated to the community sooner, this also circumvents the role of journals in filtering poor or flawed papers and highlighting relevant research \cite{67aa29563fb5476295acf440c81f2e1a}. Additionally, the uniquely multi-disciplinary nature of the scientific community's response to the pandemic has lead to pertinent research being dispersed across many open access and pre-print services - no single one of which captures the entirety of the COVID-19 literature.

These challenges have created a need and opportunity for new tools and methods to rethink the way in which researchers engage the wealth of available COVID-19 scientific literature.

COVIDScholar is an effort to address these issues by using natural language processing (NLP) techniques to aggregate, analyze, and search the COVID-19 research literature. We have developed an automated, scalable infrastructure for scraping and integrating new research as it appears, and used it to construct a targeted corpus of over 81,000 scientific papers and documents pertinent to COVID-19 from a broad range of disciplines. The search interface for this corpus, \url{https://covidscholar.org}, now serves over 2000 unique users weekly.

While a variety of other COVID-19 literature aggregation efforts exist \cite{whocovid,wang2020cord19,litcovid}, COVIDScholar differs in the breadth of literature collected. In addition to the biological and medical research collected by other large-scale aggregation efforts such as CORD-19 \cite{wang2020cord19} and LitCOVID \cite{litcovid}, COVIDScholar's collection includes the full breadth of COVID-19 research, including public health, behavioural science, physical sciences, economics, psychology, and humanities.

In this paper, we present a description of the COVIDScholar data intake pipeline and back-end infrastructure, and the NLP models used to power directed searches on the front-end search portal. We also present an analysis of the COVIDScholar corpus, and discuss trends in the dynamics of research output during the pandemic.

\section{Data Pipeline \& Infrastructure}

\begin{figure}
  \centering
  \includegraphics[width=\textwidth]{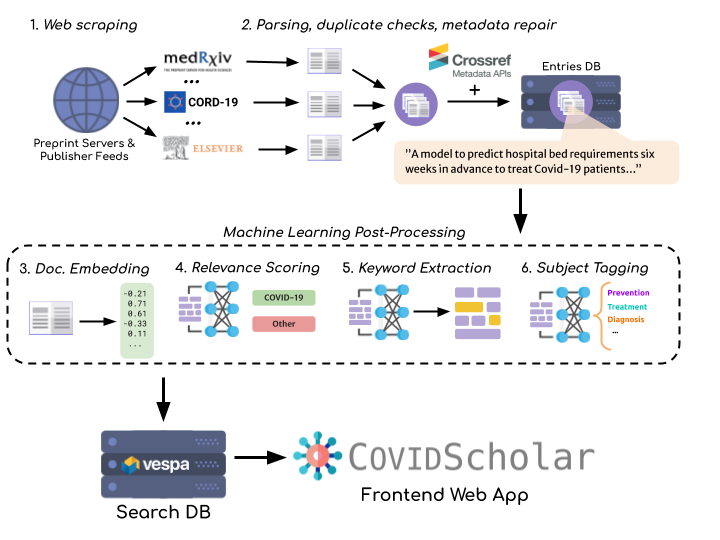}
  \caption{The data pipeline used to construct the COVIDScholar research corpus.}
\label{fig:datapipeline}
\end{figure}

At the heart of COVIDScholar is the automated data intake and processing pipeline, depicted in Fig.~\ref{fig:datapipeline}. Data sources are continually checked for new or updated papers, patents, and clinical trials, which are then parsed, cleaned, analyzed with NLP models, and made searchable on \url{https://covidscholar.org}. \footnote{The complete codebase for the data pipeline is available at \url{https://github.com/COVID-19-Text-Mining}}.

The COVIDScholar research corpus consists of research literature from 14 different open-access and pre-print services, listed in Table.~\ref{table:sources}. For each of these, a web scraper regularly checks for new documents and updates to existing ones. Missing metadata is then collected from Crossref, and citation data is collected from OpenCitations \cite{Peroni2019OpenCitationsAI}.

\begin{wraptable}{r}{0.6\linewidth}
\centering
\begin{tabularx}{\linewidth}{L | L} 
 \textbf{Source} & \textbf{COVID-19  Publications Count} \\ [0.5ex] 
 \hline\hline
preprints.org \cite{preprints} & 923 \\
osf.io \cite{osf} & 337 \\
lens.org \cite{lens} & 98 \\
SSRN \cite{ssrn} & 3491 \\
Psyarxiv \cite{rife_2016} & 691 \\
CORD-19 \cite{wang2020cord19} & 1135 \\
Dimensions.ai \cite{dimensions} & 6489 \\
Elsevier \cite{elsevier.com_2020} & 6735 \\
Chemrxiv \cite{chemrxiv}& 292 \\
LitCovid \cite{chen_allot_lu_2020}& 51807 \\
Biorxiv \cite{biorxiv}/Medrxiv \cite{rawlinson2019new} & 8832 \\
NBER.org \cite{nber} & 261 \\
COVIDScholar User Submission & 25 \\
 [1ex]
\end{tabularx}
\caption{The source of papers, patents, and clinical trials in the COVIDScholar collection, with the count of COVID-19 related publications from each source.
}
\label{table:sources}
\end{wraptable}

After collection, these publications are then parsed into a unified format, cleaned, and resolved to remove duplicates. Publications are identified as duplicates when they share any of doi (up to version number), pubmed id, or uncased title. For clinical trials without valid document identifiers, a shared title is used to identify duplicates. In cases where there are multiple versions of a single paper (most commonly, a pre-print and a published version), a combined single document is produced, whose contents are selected on a field-by-field basis using a priority system. Published versions and higher version numbers (based on doi) are given higher priority, and sources are otherwise prioritized based on the quality of their text.

In cases where full-text PDFs are available text is parsed from the document using pdfminer (for PDFs with embedded text \cite{pdfminer}) or OCR.
However, it is our experience that, in general text extracted in this manner is not of sufficient quality for to be used by the classification and relevance NLP models, and at this time is used solely for text searches.

Abstracts are classified based on their relevance to COVID-19, topic, discipline, and field. Publications are classified into 5 disciplines - Biological \& Chemical Sciences, Medical Sciences, Public Health, Physical Sciences and Humanities \& Social Sciences. A paper may belong to any number of disciplines. Each discipline is composed of 12-15 fields. The breakdown of fields by discipline is shown in the supplementary material (S.1). Publications for which an abstract cannot be found are not classified.

Keywords are also extracted from titles and abstracts using an unsupervised approach, as described in Sec.~\ref{sec:nlp}.

Our web portal, COVIDScholar.org, provides an accessible user interface to a variety of literature search tools and information retrieval algorithms tuned specifically for the needs of COVID-19 researchers. Because there still remains a great deal that we do not know about the disease, we have directed our efforts towards developing tools that can extend beyond information retrieval and aid researchers at the knowledge discovery phase as well. To do this, we have utilized new machine learning and natural language processing techniques together with proven information retrieval approaches to create the search algorithms behind COVIDScholar, which we describe in the remainder of this section. 

Machine learning algorithms can be used to identify emerging trends in the literature and correlate them with similar patterns from pre-existing research. For this reason, we chose to base our search back end on the Vespa engine \cite{vespa}, which provides a high level of performance, wide scalability, and easy integration with custom machine learning models. For example, the default search result ranking profile on COVIDScholar.org combines the BM25 relevance\cite{BM25} with a "COVID-19 relevance" score calculated by a classification model trained to predict whether a paper discusses the SARS-CoV-2 virus or COVID-19 using this approach. We observe that papers from before the COVID-19 pandemic that are related to certain viruses/diseases tend to receive high relevance scores, especially papers on the original SARS and other respiratory diseases. SARS-CoV-2 shares 79\% of its genome sequence identity with the SARS-CoV virus\cite{Lu2020}, and there are many similarities between how the two viruses enter cells, replicate, and transmit between hosts.\cite{Rabaan2020} Because the relevance classification model gives a higher score to studies on these similar diseases, search results are more likely to contain relevant information, even if it is not directly focused on COVID-19. For example, the transmembrane protease TMPRSS2 plays an important role in viral entry and spread for both SARS-CoV and SARS-CoV-2, and its inhibition is a promising avenue for treating COVID-19\cite{stopsack2020tmprss2}. A wealth of information on strategies to inhibit TMPRSS2 activity and their efficacy in blocking SARS-CoV from entering host cells was available in the early days of the COVID-19 pandemic. These studies were boosted in search results because of their higher relevance scores, thereby bringing potentially useful information to the attention of researchers more directly. In comparison, results of a Google Scholar search for "TMPRSS2" (with results containing "COVID-19" and "SARS-CoV-2" filtered out) are dominated by studies on the protease's role in various cancers.

COVIDScholar also provides tools that utilizes unsupervised document embeddings so that searches can be performed within "related documents"  to automatically link research papers together by topics, methods, drugs, and other key pieces of information. Documents are sorted by similarity via the cosine distances between unsupervised document embeddings\cite{10.5555/3044805.3045025}, which is then combined with the more overall result-ranking score mentioned above. This allows users to focus their results into a more specific domain without having to repeatedly pick and choose new search terms to add to their queries. Users can also filter all of the documents in the database by broader subjects relevant to COVID-19 (treatment, transmission, case reports, etc), which are all determined though the application of machine learning models trained on a smaller number of hand-labeled examples. All combined, these tools have allowed us to create much more targeted tools for literature search and knowledge discovery that would not be possible otherwise. 

\section{Text Analysis NLP Models} 
\label{sec:nlp}

Classification of abstracts is performed using a fine-tuned SciBERT \cite{Beltagy2019SciBERT} model. While other BERT models pre-trained on scientific text exist (e.g. BioBERT \cite{10.1093/bioinformatics/btz682}, MedBERT \cite{rasmy2020medbert}, and ClinicalBERT \cite{alsentzer-etal-2019-publicly}), we select SciBERT due to its broad, multidisciplinary training corpus, which we expect to more closely resemble the COVIDScholar corpus than those pre-trained on a single discipline. SciBERT has state-of-the-art performance on the task of paper domain classification \cite{sinha2015an}, as well as a number of biomedical domain benchmarks \cite{Yoon_2019,nye-etal-2018-corpus,10.1093/database/bay060} - the most common discipline in the COVIDScholar corpus.
A single fully-connected layer with sigmoid activation is used as a classification head, and the model is fine-tuned for 4 epochs using 2600 human-annotated abstracts \footnote{Abstracts were annotated by members of the Rapid Reviews: COVID-19 \cite{2020Rapid} editorial team.}

\begin{figure}
\centering
\includegraphics[width=\textwidth]{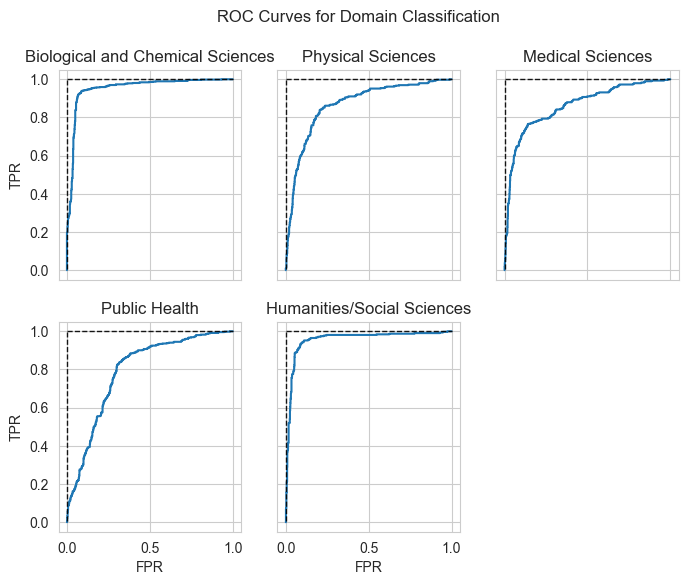}
\caption{ROC curves for discipline classification models of paper abstracts using a fine-tuned SciBERT \cite{Beltagy2019SciBERT} model adapted for classification. Training is performed using a set of 2500 human-annotated abstracts, and results shown are generated with 20-fold cross validation.}
\label{fig:disciplineroc}
\end{figure}

ROC curves for the classifier's performance for each top-level discipline using 20-fold cross-validation are shown in Fig.~\ref{fig:disciplineroc}. The classifier performs extremely well, with F1 scores above 0.73 for all disciplines. Performance metrics of the discipline classifier are displayed in Table.~\ref{table:classacc}, compared to a baseline random forest model using TF-IDF features.

\begin{table}
\centering
\begin{tabularx}{\linewidth}{L L | L | L | L | L | L} 
 \, & \, & \textbf{Biological \& Chemical Sciences} & \textbf{Medical Sciences} & \textbf{Public Health} & \textbf{Physical Sciences} & \textbf{Humanities \& Social Sciences} \\ [0.5ex] 
 \hline\hline
  \multirow{5}{2cm}{SciBERT} & F1 & 0.92 & 0.85 & 0.73 & 0.78 & 0.92 \\ 
 \cline{2-7}
 \, & Precision & 0.92 & 0.80 & 0.74 & 0.78 & 0.88\\
 \cline{2-7}
 \, & Recall & 0.92 & 0.80 & 0.75 & 0.81 & 0.92\\
 \cline{2-7}
 \, & Accuracy & 0.92 & 0.85 & 0.73 & 0.79 & 0.92 \\ 
 \hline
 \multirow{5}{2cm}{Random Forest} & F1 & 0.90 & 0.63 & 0.73 & 0.68 & 0.78 \\ 
 \cline{2-7}
 \, & Precision & 0.93 & 0.77 & 0.83 & 0.81 & 0.89\\
 \cline{2-7}
 \, & Recall & 0.89 & 0.55 & 0.67 & 0.59 & 0.73\\
 \cline{2-7}
 \, & Accuracy & 0.92 & 0.84 & 0.81 & 0.83 & 0.90 \\ 
 [1ex]
\end{tabularx}
\caption{Scoring metrics of SciBERT \cite{Beltagy2019SciBERT} and baseline random forest discipline classification models. Models were evaluated using 10-fold cross-validation on 2600 labeled abstracts. Input features to the random forest model generated using TF-IDF.
}
\label{table:classacc}
\end{table}

On three disciplines (Medical Sciences, Physical Sciences, and Humanities \& Social Sciences) the SciBERT-based discipline classifier offers a significant performance advantage over the baseline random forest/TF-IDF model, with F1 scores which are between 0.1 and 0.14 higher. These are the broadest disciplines, encompassing multiple disparate fields. The large variability of subjects within these domains may account for the inability of TF-IDF-based models to classify them well.

For the remaining two disciplines, Biological \& Chemical Sciences and Public Health, the F1 scores are similar between SciBERT and the baseline model. In the case of Biological \& Chemical Sciences, this may be explained by relatively distinctive vocabulary and narrow subjects within the discipline. Public Health was observed to have the largest inter-annotator disagreement, leading to a lower performance by the classifier.

It is also of note in each case that while precision is broadly similar between the two models, the baseline model exhibits significantly lower recall. This may be due to unbalanced training data - no single discipline accounts for more than 33\% of the total corpus. For search applications, often a relatively small number of documents is relevant to each query. In this case, a high recall is more desirable than a high precision - in practice, the performance gap between the two models is larger than indicated by relative F1 scores.

On the task of binary classification as related to COVID-19, our current models perform similarly well, achieving an F1 score of 0.98. While the binary classification task is significantly simpler from an NLP perspective - the majority of related papers contain "COVID-19" or some synonym - this still represents a significant performance improvement over the baseline model, which achieves an F1-score of 0.90. Given the relative simplicity of this task, in cases where an abstract is absent we classify it as related to COVID-19 based on the title.

For the task of unsupervised keyword extraction, 63 abstracts were annotated by humans, and two statistical methods, TextRank \cite{mihalcea-tarau-2004-textrank}and TF-IDF \cite{SALTON1988513}, and two graph-based models, RaKUn \cite{Skrlj2019RaKUnRK} and Yake \cite{inproceedings}, were tested. Models were evaluated for overlap between human-annotated keywords and extracted keywords, and results are shown in Table.~\ref{table:keywords}. Note that due to the inherent subjectivity of the keyword extraction task that scores are relatively low - the best performing model, RaKUn has an F1 score of only 0.2. However, the quality of extracted keywords from this model was deemed reasonable for display on the search portal after manual inspection.

\begin{wraptable}{l}{0.6\linewidth}
\centering
\begin{tabular}{c c c c} 
 Model & Precision & Recall & F1 \\ [0.5ex] 
 \hline\hline
 RaKUn & 0.17 & 0.33 & 0.2 \\ 
 \hline
 Yake & 0.11 & 0.45 & 0.15 \\
 \hline
 TextRank & 0.06 & 0.36 & 0.09 \\
 \hline
 TF-IDF & 0.10 & 0.09 & 0.08 \\ [1ex]
\end{tabular}
\caption{Precision, recall, and F1 scores for 4 unsupervised keywords extractors, RaKUn\cite{Skrlj2019RaKUnRK}, Yake\cite{inproceedings}, TextRank\cite{mihalcea-tarau-2004-textrank}, and TF-IDF\cite{SALTON1988513}. Output from keyword extractors was compared to 63 abstracts with human-annotated keywords.
}
\label{table:keywords}
\end{wraptable}

\begin{figure}[h!]
\centering
\includegraphics[width=\linewidth]{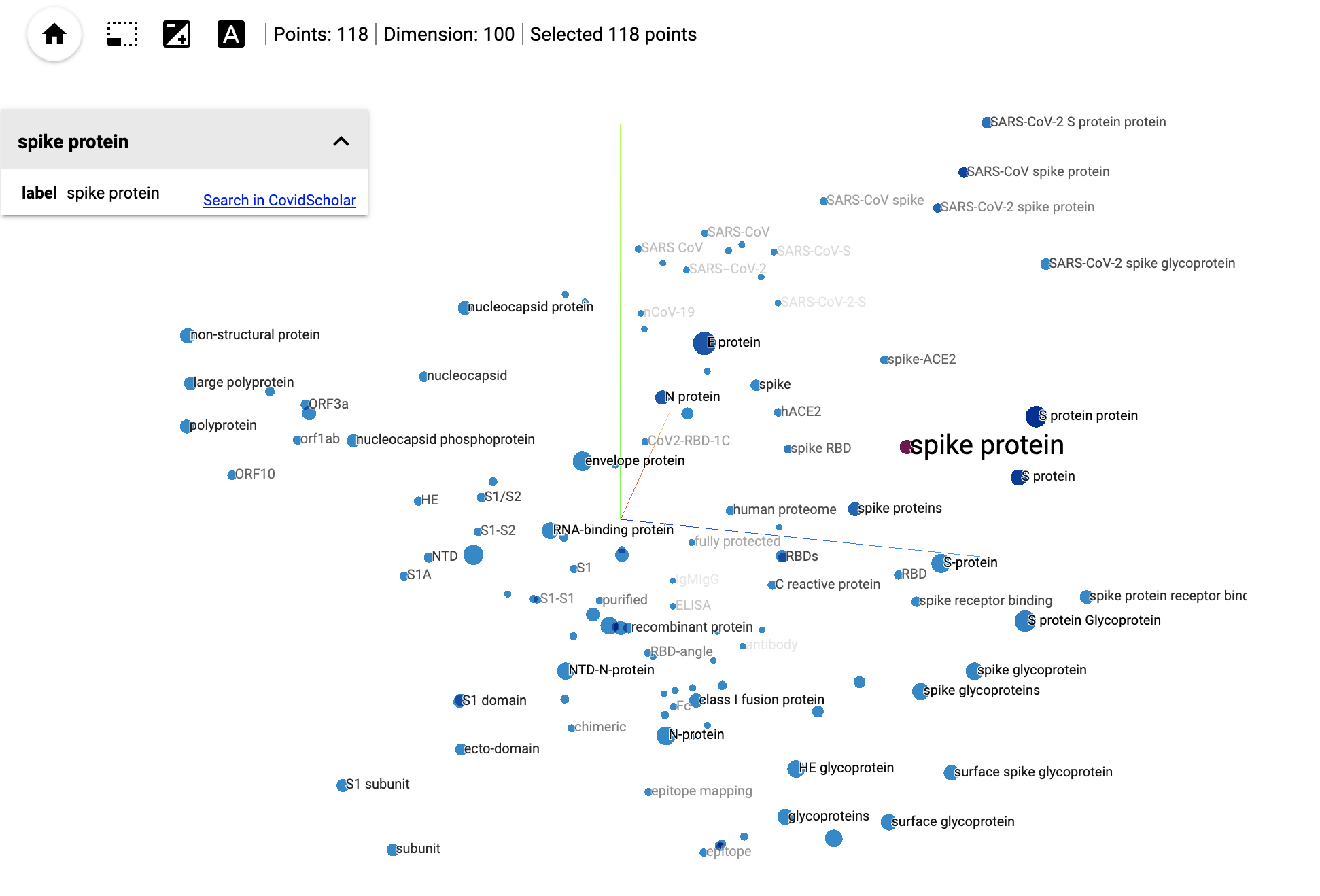}
\caption{A screenshot of the embedding projector visualizing tokens similar to "spike protein", using FastText\cite{bojanowski2016enriching} embeddings trained on the COVIDScholar corpus.}
\label{fig:embeddingprojector}
\end{figure}

To better visualize the embedding of COVID-19-related phrases and find latent relationship between biomedical terms, we designed a tool based on Embedding Projector\cite{smilkov2016embedding}. A screenshot of the tool is shown in Fig.~\ref{fig:embeddingprojector}

We utilize FastText\cite{bojanowski2016enriching} embeddings for the embedding projector, with an embedding dimension of 100. Embeddings are trained on the abstracts of all papers which have been classified as relevant to COVID-19. 

For the purpose of visualization, embeddings must be projected to a lower dimensional space (2D or 3D). The dimensionality reduction technique used here includes principal component analysis (PCA), uniform manifold approximation and projection (UMAP) and t-distributed stochastic neighbor embedding (t-SNE). Users can set various parameters and do the dimension reduction via an interactive page. They can also load and visualize the cached result on the server with default parameters.


Cosine distance is used to measure the similarity between phrases. If the cosine distance between two phrases is quite small, they are likely to have similar meaning.

\[
\text{Cosine Distance}(p_1, p_2) = \frac{\textit{\textbf{Emb}}(p_1)\cdot \textit{\textbf{Emb}}(p_2)}{\|\textit{\textbf{Emb}}(p_1)\|\|\textit{\textbf{Emb}}(p_2)\|}
\]

$p_1, p_2$ represent two phrases, $\textit{\textbf{Emb}}$ maps phrases to their embedded representation in the learned semantic space.

\section{COVIDScholar Corpus Analysis}

\subsection{Corpus Breakdown}
\label{sec:corpus}

As of October 2020, the COVIDScholar corpus consists of 150,113 total documents, of which 143,887 are papers. The remainder is composed of 3306 patents, 1712 clinical trials, 1025 book chapters, and 183 datasets. Of the papers, 81,106 are classified as related to COVID-19,\footnote{Papers marked not relevant to COVID-19 are a combination of papers on related diseases, such as SARS and MERS, and with no relation to COVID-19.} and are approximately equally split between preprints and published papers - 44\% pre-prints, 56\% published. A breakdown by discipline of the COVID-19 relevant papers is shown in Table.~\ref{table:discipliecount}. As may be expected, Public Health and Biological \& Chemical Sciences are the most represented disciplines, with respectively 56\% and 42\% of the corpus tagged as members of these disciplines. Overlap between these two disciplines is relatively small ---only 3295 papers are classified as belonging to both Public Health and Biological \& Chemical Sciences---, and so the vast majority of the corpus, 50,787 papers, belongs to one of the two.

\begin{table}
\centering
\begin{tabularx}{\linewidth}{L | L | L} 
 \textbf{Discipline} & \textbf{Paper Count} & \textbf{Fraction of Total} \\ [0.5ex] 
 \hline\hline
Biological \& Chemical Sciences & 23227 & 0.42\\
\hline
Humanities \& Social Sciences & 17464 & 0.31\\
\hline
Medical Sciences & 21023 & 0.38\\
\hline
Physical Sciences & 17214 & 0.31 \\
\hline
Public Health & 30855 & 0.56\\
 [1ex]
\end{tabularx}
\caption{The number of papers and fraction of total COVID-19 related papers in the COVIDScholar corpus for each discipline. Only papers with abstracts are classified and included in final count. Note that a given paper may have any number of discipline labels.
}
\label{table:discipliecount}
\end{table}

\subsection{Research Trends}

\begin{figure}
  \centering
  \includegraphics[width=\textwidth]{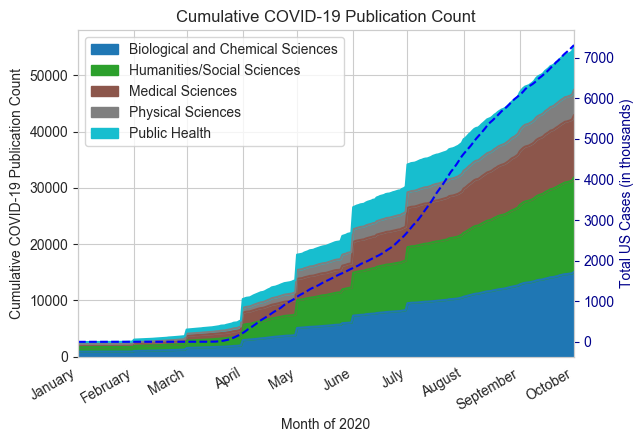}
  \caption{Cumulative count by primary discipline of COVID-19 papers in the COVIDScholar database, and total number of reported US COVID-19 cases during the first 10 months of 2020. Papers are categorized by the classification model described in Sec.~\ref{sec:nlp}, and assigned to the discipline with highest predicted likelihood. Case data from The New York Times, based on reports from state and local health agencies. Note that only those papers with abstracts available are classified, and so the publication is somewhat lower than the total from Sec.~\ref{sec:corpus}.}
\label{fig:pubcount}
\end{figure}

The cumulative count of COVID-19 papers in the COVIDScholar collection over the first 10 months of 2020 is shown in Fig.~\ref{fig:pubcount}. Papers are categorized by the discipline with highest predicted likelihood using the fine-tuned SciBERT model described in Sec.~\ref{sec:nlp}. Note that papers for which the day of publication is unknown are assigned to the first of the month, causing the step-like features visible at the beginning of each month. The total number of reported US COVID-19 cases is also plotted. Data on cases is from The New York Times, based on reports from state and local health agencies (\url{https://www.nytimes.com/interactive/2020/us/coronavirus-us-cases.html}).

The rate at which publications emerged in all disciplines shows a steep increase through the early months of 2020. Between the declaration of a Public Health Emergency of International Concern\cite{whodeclar} by the World Health Organization in January 2020 and April 2020, the rate of new publications approximately tripled each month, from just 91 papers in January to 7135 in April. From May onwards, the rate stabilized at approximately 8000 papers a month.

Given the lag between research and publication, it therefore seems that by April 2020 the COVID-19 research effort had already reached full capacity, before the US case count began to dramatically rise in the Summer. The US government passed two stimulus bills, each with over \$1 billion in funding allocated  for coronavirus research on March 5th \cite{march5stim} and March 27th \cite{march27stim}. The data suggests that any increase in rate of research associated with these had already fully manifested itself within 2 months of their passing, demonstrating the rapidity of the scientific community's COVID-19 response. Other notable events within this timeframe include the declaration of global pandemic by the WHO on March 11 \cite{whopandemic}.

\begin{figure}
  \centering
  \includegraphics[width=\textwidth]{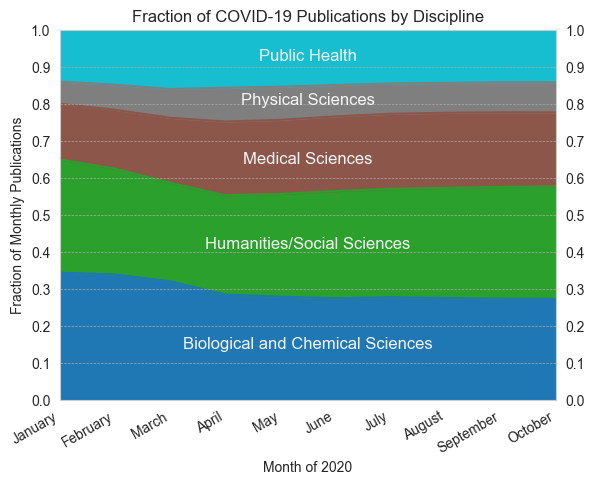}
  \caption{Fraction of total COVID-19 papers by primary discipline. Fractions are calculated based on total over previous calendar month. Papers are categorized by the classification model described in Sec.~\ref{sec:nlp}, and assigned to the discipline with highest predicted likelihood.}
\label{fig:fracpubcount}
\end{figure}

A breakdown of research by discipline over the course of 2020 is shown in Fig.~\ref{fig:fracpubcount}, which depicts the fraction of monthly COVID-19 publications primarily associated with each discipline. From January - April, the relative popularity of discipline showed some shifts. While Biological and Chemical Sciences comprised 45\% of the total corpus in January, by April that had decreased to 28\%. This is largely accounted for by an increase in papers from Physical and Medical Sciences - over the same period the fraction of papers from Medical Sciences increased from 15\% to 20\% of the total, and Physical Sciences from 5\% to 8\%. By April, the fraction of the corpus from each discipline seems to have stabilized, with fluctuations of relative fractions of under 1\%. This is further support for the evidence in Fig.~\ref{fig:pubcount} that research output had already reached its maximum rate by April/May - this seems to hold true on a discipline-by-discipline basis also.

\begin{figure}
  \centering
  \includegraphics[width=\textwidth]{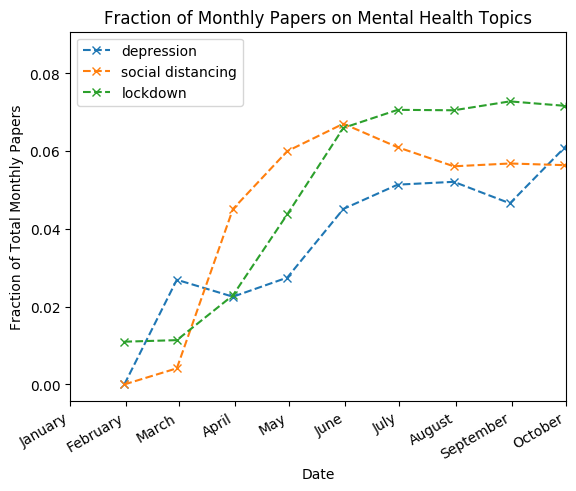}
  \caption{Fraction of COVID-19 literature on mental health- and lockdown- related topics on a monthly basis.}
\label{fig:mentalhealth}
\end{figure}

We investigate this increase in Fig.~\ref{fig:mentalhealth}, where we have plotted the fraction of total monthly papers on selected mental health- and lockdown- related topics. Over the April-June period, there is a clear increase in research related to psychological impacts of lockdown and social distancing, accounting for 6-8\% of total monthly papers. Between March and April, many countries and territories instituted lockdown orders, and by April, over half of the world's population was under either compulsory or recommended shelter-in-place orders \cite{lockdown}. The corresponding emergence of a robust literature on psychological impacts associated with this is the major driving force behind the increase in COVID-19 literature from Humanities \& Social Sciences.

\section{Summary and Future Work}

We have developed and implemented a scalable research aggregation, analysis, and dissemination infrastructure, and created a targeted corpus of over 81,000 COVID-19 relevant research documents. The associated search portal, \url{https://covidscholar.org}, serves over 2000 weekly scientific users.

While the large amount of open data and enormous scientific interest in COVID-19 have made it an ideal use-case, the infrastructure is domain-agnostic, and presents a blueprint for future large-scale scientific literature aggregation efforts.

While to-date the COVIDScholar research corpus has primarily been used for front-end user search, it provides a rich opportunity for NLP analysis. Recent work \cite{tshitoyan2019} has highlighted the ability of NLP to discover latent knowledge from unstructured scientific text, utilizing information from thousands of research papers. We are now moving to employ similar techniques here, applied to such problems as drug re-purposing and predicting protein-protein interactions.

\section{Acknowledgements}

Portions of this work were supported by the C3.ai Digital Transformation Institute and the Laboratory Directed Research and Development Program of Lawrence Berkeley National Laboratory under U.S. Department of Energy Contract No. DE-AC02-05CH11231.

The text corpus analysis and development of machine learning algorithms were supported by the DOE Office of Science through the National Virtual Biotechnology Laboratory, a consortium of DOE national laboratories focused on response to COVID-19, with funding provided by the Coronavirus CARES Act.

This research used resources of the National Energy Research Scientific Computing Center (NERSC), a U.S. Department of Energy Office of Science User Facility operated under Contract No. DE-AC02-05CH11231.

We are thankful to the editorial team of Rapid Reviews: COVID-19 for their assistance in annotating text.

\printbibliography
\end{document}


\title{Supplementary Material for COVIDScholar: Analyzing the COVID-19 research literature with NLP}
\section{Discipline and Field Classifications}

\begin{table}
\centering
\begin{tabularx}{\linewidth}{L | L | L | L | L} 
 \textbf{Biological \& Chemical Sciences} & \textbf{Medical Sciences} & \textbf{Public Health} & \textbf{Physical Sciences} & \textbf{Humanities/Social Sciences} \\ [0.5ex] 
 \hline\hline
 Virology & Pathophysiology & Epidemiology & Engineering & Education \\ 
 \hline
 Immunology & Genetics & Biostatistics/ Disease Modelling & Physics & Psychology\\
 \hline
 Genetics/ Genomics/ Epigenetics & Clinical Management & Health Policy & Data Science & Anthropology\\
 \hline
 Biomedical Engineering/ Biotechnology & Infectious Disease & Nutrition/ Food Science & Mathematics  Journalism and Communications & Agriculture \& Resource Management \\ 
 \hline
 Vaccinology & Rheumatology & Community Health & Computational Sciences & Law \& Ethics\\
 \hline
 Toxicology & Pulmonology/ Critical Care & Environmental Health & Materials Science & Economics \\
 \hline
 Molecular \& Cell Biology & Geriatrics & Occupational Health & Chemistry \& Chemical Engineering & Political Science\\
 \hline
 Integrative Biology & Emergency Medicine & Implementation Science & Statistics & International Relations \\
 \hline
 Ecology, Evolution, Biodiversity & Cardiology & Health Diplomacy & Climate Science & Design/ Arts/ Literature/ Music \\
 \hline
 Plant Biology & Pediatrics & Health Communications & Earth Sciences/ Geosciences & Sociology \\
 \hline
 Biochemistry & Obstetrics \& Gynecology & Behavioural Health & Zoology & Business\\
 \hline
 Pathology/ Laboratory Medicine & Nephrology & Non-Communicable Diseases & Botany & Gender, Sexuality \& Women's Studies \\
 \hline
 Zoonotic Diseases & Clinical Trials & \, & \, & History \\
 \hline
 \, & Primary Care & \, & \, & Area Studies\\
 \hline
 \, & Pharmacology & \, & \, & Philosophy\\
 [1ex]
\end{tabularx}
\caption{The 5 top-level disciplines (boldface) and corresponding composite fields into which COVIDScholar's text corpus is classified.
}
\label{table:disciplines}
\end{table}